\documentstyle[aps,prb,epsf]{revtex}
\begin{document}

\title{Josephson array of mesoscopic objects.  Modulation of system
properties through the chemical potential}
\author {A. I. Belousov, S. A. Verzakov, and Yu. E. Lozovik\cite{A}}
\address{Institute of Spectroscopy, Russian Academy of Sciences,\\
142092 Troitsk, Moscow Region, Moscow}
\maketitle

Zh. \'{E}ksp. Teor. Fiz. \bf{114}\rm, 591--604 (August 1998)

\begin{abstract}
The phase diagram of a two-dimensional Josephson array of mesoscopic
objects (superconducting granules, superfluid helium in a porous
medium,
traps with Bose-condensed atoms, etc.) is examined.  Quantum
fluctuations
in both the modulus and phase of the superconducting order
parameter are taken into account within a lattice boson Hubbard
model.  Modulating the average occupation number $n_0$ of the sites
in the system (the ``number of Cooper pairs'' per granule, the number
of atoms in a
trap, etc.) leads to changes in the state of the array, and the character of
these changes depends significantly on the region of the phase diagram
being examined.  In the region where there are large quantum
fluctuations
in the phase of the superconducting order parameter, variation of the
chemical potential causes oscillations with alternating superconducting
(superfluid) and normal states of the array.  On the other hand, in the
region where the bosons interact weakly, the properties of the system
depend monotonically on $n_0$.  Lowering the temperature and
increasing
the particle interaction force lead to a reduction in the width of the
region
of variation in $n_0$ within which the system properties depend weakly
on the average occupation number.  The phase diagram of the array is
obtained by mapping this quantum system onto a classical
two-dimensional $XY$ model with a renormalized Josephson coupling
constant and is consistent with our quantum Path-Integral Monte Carlo
calculations.
\end{abstract}

\renewcommand{\thesection}{\arabic{section}}
\section{INTRODUCTION}

Advances in microlithographic techniques have made it possible to create
regular arrays of mesoscopic Josephson objects.  These systems, which
are under active experimental and theoretical study, include superfluid
helium in porous media\cite{1)}$^{)}$ (see Ref. \onlinecite{2} and the
literature
cited there), lattices of mesoscopic Josephson contacts,\cite{3,4} and
ultrasmall superconducting granules.\cite{5,6}  One interesting physical
realization of a Josephson array consists of Josephson junctions in
a structure with
superfluid $^3$He created by means of lithography.\cite{7}  Major
advances
in experiments with Bose condensates of atoms cooled by laser
irradiation followed by evaporation\cite{8,9,10} suggest that it may be
possible to fabricate
a Josephson array of close magneto-optical traps with Bose-condensed
atoms\cite{2)}$^{)}$ or with clusters of Bose-condensed atoms cooled
and localized at the nodes of a system of standing electromagnetic
waves.  Finally,
another remarkable realization of a Josephson array might be a system
of
Josephson coupled ``lakes'' of Bose-condensed excitons in single or
double
quantum wells located at the minima of the random field created by the
roughness of the well surfaces, i.e., in ``natural'' quantum dots,\cite{12}
or in an array of artificial quantum dots.

For concreteness, the following discussion considers the example of a
system of superconducting mesoscopic granules or Josephson junctions,
but the results apply to all the systems mentioned above.  We
consider a regular array of mesoscopic granules situated on a conducting
substrate and separated from it by a thin dielectric layer.  A voltage
applied to the conducting substrate serves as the chemical potential of
the
Cooper pairs, which determines the average occupation number $n_0$ of
the
granules in the system.\cite{13,14}  For example, in the case of
superfluid
helium in a porous medium, the chemical potential of the atoms can be
varied (because of the contribution of the van der Waals interaction) by
changing the thickness of the layer of adsorbed helium.\cite{15,16}  The
state of an array of mesoscopic traps with cooled atoms can be
controlled
by changing the average number of atoms in the system (by, for
example,
capturing them from an external flow).

The systems being examined here, in which the character of the
processes
taking place is determined by the boson degrees of freedom, are
conveniently described by a lattice boson Hubbard model\cite{17,18}
with the Hamiltonian
\begin{equation}
\hat H =
\frac{t}{2} \sum\limits_{<i,j>}
\left( 2 a{^\dagger}_i a_i -  a{^\dagger}_i a_j - a{^\dagger}_j a_i \right) +
\frac{U}{2} \sum\limits_{i}
\left( a{^\dagger}_i a_i \right)^2 -
\mu \sum\limits_{i} a{^\dagger}_i a_i
\end{equation}
In this model site $i$ corresponds to one superconducting granule or
pore
with helium, to a single trap with a Bose condensate, etc.  The operators
$a_i^{\dagger }$ ($a_i$) are the Bose creation (annihilation) operators
of an ``effective'' boson at site $i=\overline{1,N^2}$ of an $N \times
N$
lattice.  The first term in the Hamiltonian takes into account the kinetic
energy
of the particles, which corresponds to the Josephson tunneling energy
and is
described by the parameter $t$.  The second term in Eq. (1) describes
the
interaction of effective bosons at a granule with a characteristic energy
$U_i>0$.

The model (1) is interesting in that it can be used to study the properties
of arrays of mesoscopic structures in which the relative fluctuations in
the modulus of the superconducting order parameter are large.  In
this
regard, we note that the quantum $XY$ model is justified only
when these fluctuations are small,\cite{14} i.e., in the case of arrays of
macroscopic granules.

A lattice system (1) with Mott-insulator and superconducting phases at
$T=0$ has been studied both analytically\cite{17,18,19} and by
computer
simulation.\cite{20,21}  In this paper we shall be interested in the
properties of the system (1) at finite temperatures, digressing from the
interesting phase transitions at $T=0$.\cite{22}  In our earlier
papers,\cite{23} the investigation was limited to the case of integer
(commensurable) populating, in which the average number of bosons
per
site (granule), $n_0= \langle a_i^{\dagger }a_i \rangle $, is a whole
number.
At
$T=0$, adding even a single particle to an arbitrarily large system
changes
its properties in a fundamental way.  Specifically, a system with a
noninteger average number of bosons per granule remains
superconducting
for arbitrary values of $U/t$ for the interaction between the
particles.\cite{17,18,19,20,21,22}  It is evident that this surprising
behavior should be present even in the limit of $T=0$.  In fact, as will
be
shown below, at finite temperatures the properties of the system vary
little within an interval $n_0=k \pm \delta n_0$ about an integer
population, whose width $2 \delta n_0$ decreases as the temperature is
lowered.

The purpose of this paper is to study the changes in the character of the
ordering in an array of granules as the substrate voltage (chemical
potential of the effective bosons) is varied.  Here we shall not use the
simplifying assumption of small relative fluctuations in the modulus of
the superconducting order parameter.  The results given below
correspond to an array of mesoscopic objects for which the
root-mean-square fluctuations in the number of particles are comparable
to their
average number.  In Sec. 2 we present results from mean-field
calculations.  The method used there corresponds to
mapping the initial boson model (1) onto an effective classical $XY$
model
with a renormalized Josephson coupling constant.  To refine the results
of
the analytical calculations, we use the quantum Path-Integral Monte
Carlo method (see Sec. 3).  A discussion and
comparison of the results in Sec. 4 completes the presentation.

\section{BOSON HUBBARD MODEL IN THE MEAN-FIELD
APPROXIMATION}

A qualitative approximation for the phase diagram of the model (1) can
be
obtained using an approach described previously.\cite{19,23,24}  In
terms
of this model, the boundary of the ordered state is located where the
local
density of the superconducting component vanishes in the
effective
functional describing long-wavelength excitations of the system.  The
latter
can be obtained in the usual way using the Hubbard--Stratonovich
transform\cite{25} followed by an expansion of the effective functional
in
components of the fluctuating field.\cite{19,24}  The condition that the
local density of the superconducting component vanish for the
system
described by the Hamiltonian (1) yields an equation for determining the
boundary of the ordered state:
\begin{equation}
\tilde q^2 = \frac{
4 \sum\limits_{n=0}^{\infty}
\left[ \exp{\left\{ -0.5\tilde q^2 (n-\tilde \eta)^2/\tilde T \right\} }
-
\exp{ \left\{-0.5\tilde q^2 (n+1-\tilde \eta)^2/\tilde T \right\}}
\right] (n+1) / (2(n-\tilde \eta)+1)
}
{
\sum\limits_{n=0}^{\infty} n
\exp{ \left\{ -0.5\tilde q^2 (n-\tilde \eta)^2/\tilde T \right\} }
}
\end{equation}
Here we have used the following independent dimensionless parameters,
which specify the state of the system:
$$
\tilde q = \sqrt{U/t}, \qquad
\tilde T = \frac{k_b T}{t}, \qquad
\tilde \eta = \frac{\mu}{U} - \frac{2 t}{U}
$$
The average number of particles in the granules of the system in the
disordered state is given by the equation
\begin{equation}
n_0 = \frac{
\sum\limits_{n=0}^{\infty} n
\exp{ \left\{ -0.5\tilde q^2 (n-\tilde \eta)^2/\tilde T \right\} }
}{
\sum\limits_{n=0}^{\infty}
\exp{ \left\{ -0.5\tilde q^2 (n-\tilde \eta)^2/\tilde T \right\} }
}
\end{equation}

The solid curves in Fig. \ref{fig1}a represent the phase diagram of the
system (1) in the variables $\{\sqrt{U/t}, \mu /U \}$ obtained by
solving Eq. (2)
for different temperatures $k_BT/t$.  The region of low values of $U/t$
(low
particle interaction energies) corresponds to the
superconducting (S) state of the system.

\begin{figure}
\epsfxsize = 14truecm
\epsffile{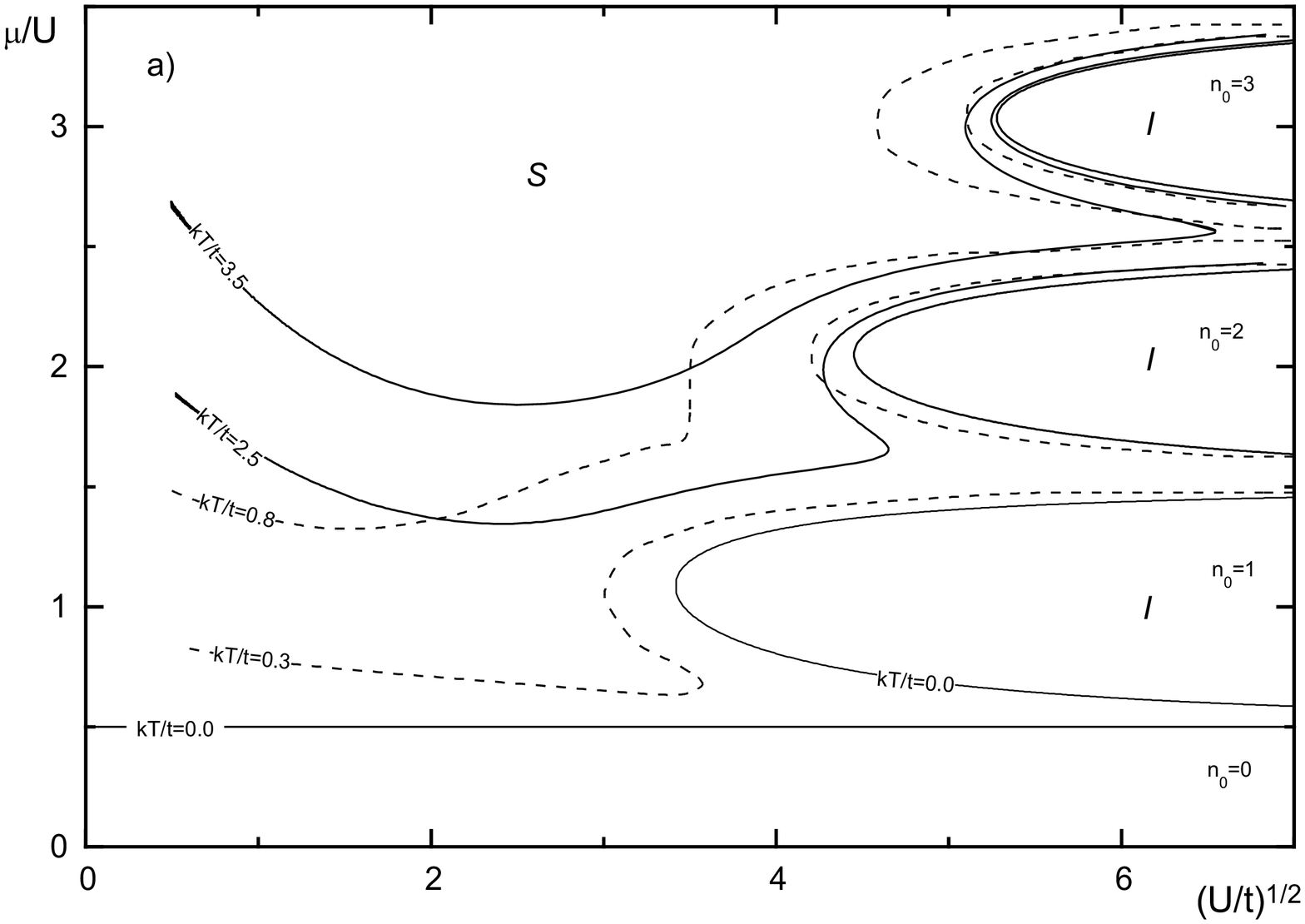}
\end{figure}
\begin{figure}
\epsfxsize = 14truecm
\epsffile{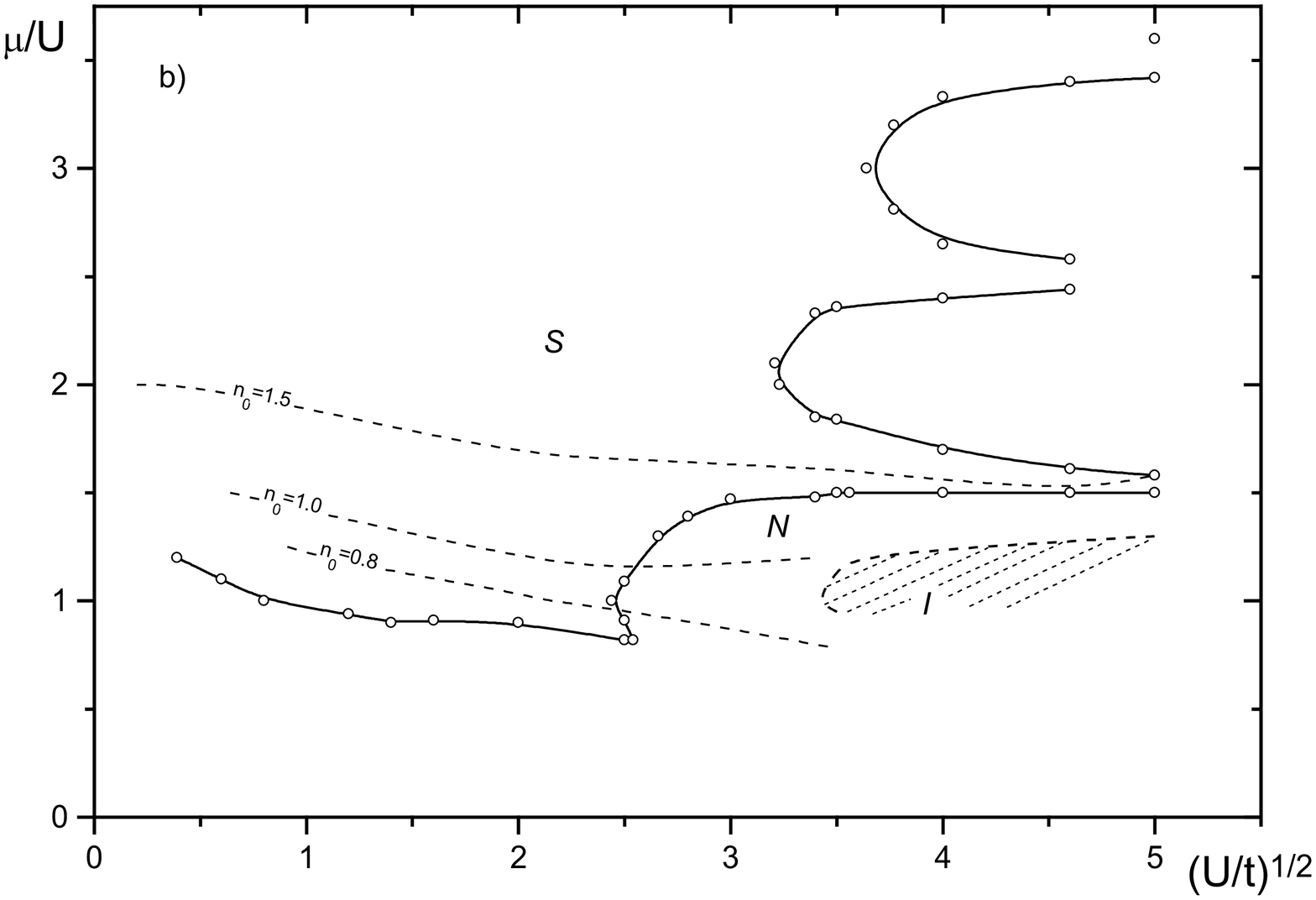}
\caption[]{Phase diagram for the Hubbard model (1) in the coordinates
$\{ \sqrt{U/t}, \mu
/U \}$.  S --- the superconducting state; N --- the normal (metallic)
state; I --- Mott insulator (hatched region in Fig. b). (a)
Mean-field calculation.  The solid curves were
obtained by solving Eq. (2) and correspond to vanishing of the local
density
of the superfluid component.  The Kosterlitz--Thouless topological
transition (6) takes place on the dashed curves.  (b) Monte Carlo
calculations for $k_BT/t=0.8$.}
\label{fig1}
\end{figure}

\begin{figure}
\epsfxsize = 14truecm
\epsffile{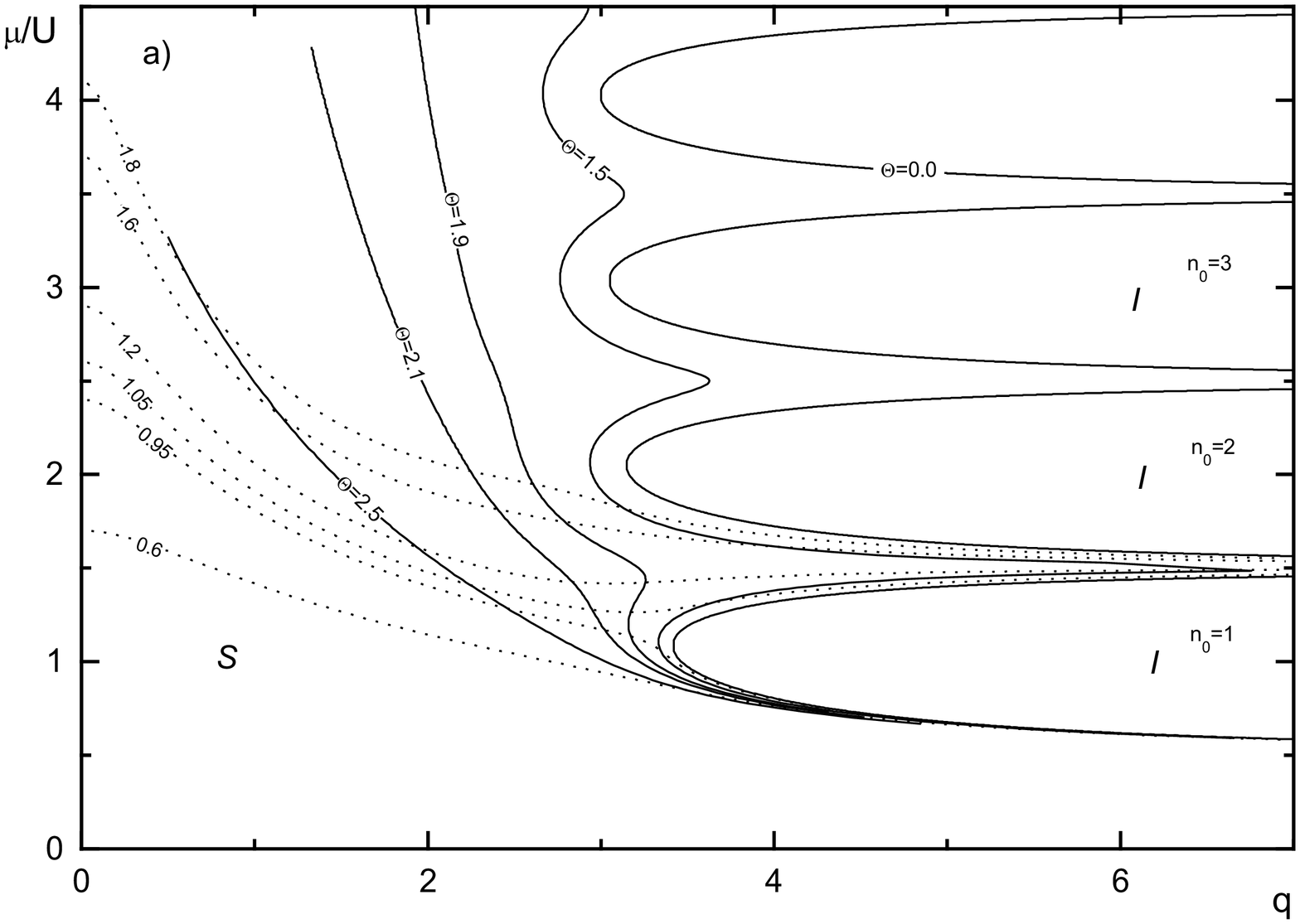}
\end{figure}
\begin{figure}
\epsfxsize = 14truecm
\epsffile{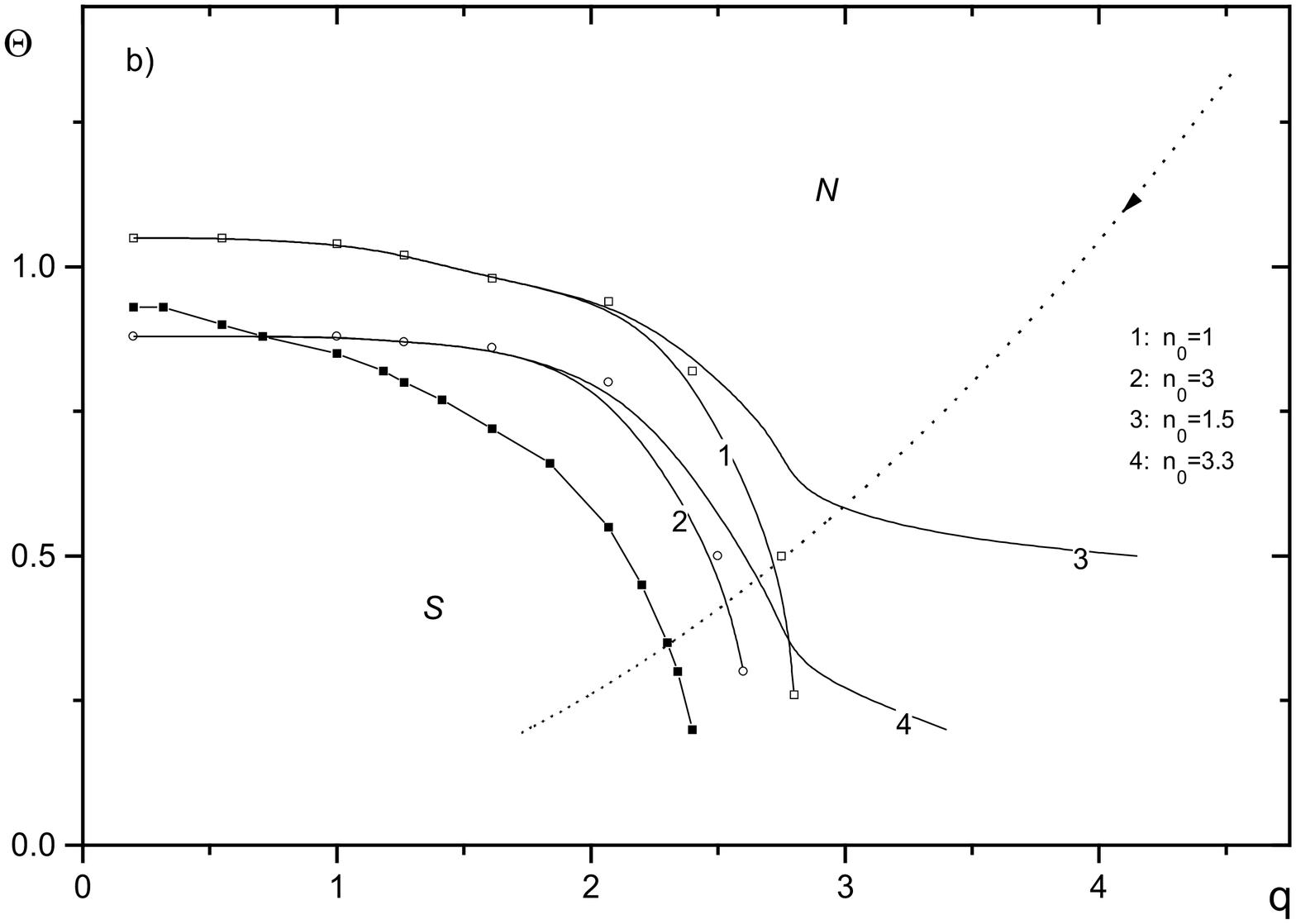}
\caption[]{Phase diagram for the Hubbard model (1) in the coordinates
$\{ q, \mu /U \}$
and
$\{ q, \Theta \}$.  (a) Mean-field theory calculation.
The
solid curves were obtained by solving Eq. (2).  The dotted curves
correspond to the joint solution of the system of Eqs. (2) and (3).  (b)
Monte Carlo calculations.  The filled squares are the phase diagram for
the
$2+1$-dimensional $XY$ model (for integer values of $n=k \gg
1$).\cite{34}
The
system with $\{ U/t, k_B T/t \} =\{3.5, 0.8 \}$ and a variable chemical
potential $\mu /U$ moves along the dotted curve (see Figs. \ref{fig4}
and \ref{fig5}
below).}
\label{fig2}
\end{figure}

As $k_BT /t \rightarrow 0$, the disordered state of the system
corresponds to an integer average number of particles per granule,
$n_0=k$, and determines the domain of existence of the Mott insulator
(I).\cite{17}  In this limit, the superfluid state of the system corresponds
to the case of incommensurate populations, i.e., to a noninteger average
number of particles per granule.  Figure \ref{fig1}a shows that for
$T=0$
and half-integer values of the chemical potential $\mu =0.5+k$, a
superconducting state exists for arbitrarily strong interparticle
interactions, in accord with earlier works.\cite{17,18}

At finite temperatures an increase in the boson interaction force leads
the system into a disordered state for arbitrary values of the chemical
potential (see Fig. \ref{fig1}a).  It is also clear from the figure that as
the temperature is increased, the domain of existence of the ordered
state is shifted toward higher values of the chemical potential.

The transition to the case of a system of macroscopic granules
corresponds to increasing the particle density $n_0$ and reducing the
role of the fluctuations in the modulus of the order parameter.  It is
most convenient to follow the changes in system (1) with increasing
$n_0$ in the $\{ q , \Theta \}$ plane, where we use the dimensionless
temperature $\Theta = k_B T/tn_0$ and the quantum
parameter $q= \sqrt{U/tn_0}$, which are the control parameters that
also determine the state of the quantum $XY$ model.  The
corresponding phase diagram is shown in Fig. \ref{fig2}a.  As can be
seen
from this figure, for any values of $U$ the estimate of the
boundary of the ordered state in the Hubbard model according to
mean-field theory lies above the corresponding limit in the $XY$
model
and approaches it as the average population $n_0$ of the lattice sites
increases.  Our calculations confirm that the phase diagram is
periodic in the parameter $\mu /U$ when $n_0 \gg 1$.\cite{14}

The dotted lines in Fig. \ref{fig2}a comprise a family of curves, at
whose points a system with density $n_0$ becomes disordered.  The
points where they intersect the $\Theta ={\rm const}$ lines determine
the phase
diagram of a system in the $\{ q, \Theta \}$ plane corresponding to
$n_0$
particles per granule.  A similar analysis shows that for an
incommensurate boson density at low temperatures (see below), an
ordered (superconducting) state of an array exists for
arbitrarily large values of
$q$, i.e., for arbitrarily large quantum fluctuations in the
phase of the superconducting order parameter in terms of the
quantum $XY$ model.

This approach only yields a qualitative estimate of the characteristic
features of the phase diagram of this system.  A comparison with the
results of a numerical simulation (see below) shows that Eqs. (2) and (3)
give a greatly overestimated value for the disorder temperature $\Theta
_c(q; \mu /U)$.  In order to obtain more accurate quantitative estimates,
it is necessary to determine the temperature at which the global (rather
than
local, as in the method described above) superfluid density of the array
vanishes.  This temperature can be estimated as the Kosterlitz--Thouless
temperature for the topological phase transition according to the classical
$XY$ model, onto which the initial system is mapped by expanding the
effective Ginzburg--Landau functional for weak fluctuations in the
phase of the order parameter.

Using the approach in Refs. \onlinecite{19,24}, and \onlinecite{26}
it is easy to show
that the effective action of the classical two-dimensional $XY$
model sought is given by
\begin{equation}
S(\{\varphi_{\vec k}\}) = \frac{J_{XY}}{2}\sum\limits_{\vec k}
|\vec k|^2 \varphi_{\vec k} \varphi_{- \vec k} \approx
J_{XY}\sum\limits_{<i,j>}\left( 1 -
\cos{(\varphi_i - \varphi_j)} \right),
\nonumber
\\
J_{XY}(\mu/U;t/U;k_bT/t) = \frac{t \Delta^2}{4},
\end{equation}
where $J_{XY}$ is the coupling constant for the effective $XY$ model,
which depends on the dimensionless parameters $\mu /U$, $t/U$, and
$k_B
T/t$.  The local superfluid density $\Delta ^2/4$ of the system is
described by the relation
\begin{equation}
\Delta = \frac{tr\left\{ (\hat a^{\dagger} + \hat a) e^{-\beta \hat H_{mf}}\right\}}
{tr\left\{ e^{-\beta \hat H_{mf}} \right\}},
\\
\hat H_{mf} = \frac{U}{2} \hat n^2 + (2t - \mu) \hat n  -
t \Delta  (\hat a^{\dagger} + \hat a)
\end{equation}
A plot of the Kosterlitz--Thouless phase-transition temperatures in the
effective $XY$ model\cite{27} specifies the boundary sought of the
superfluid state for this array of granules:
\begin{equation}
k_b T^c = 0.98 J_{XY}(\mu/U;t/U;k_bT^c /t)
\end{equation}

Estimates given by Eqs. (4) and (5) are shown as dashed curves in
Fig.
\ref{fig1}a.  Note that, although the two approaches discussed above
yield a similar
qualitative behavior of the boundary of the ordered phase, the
temperature
of the topological phase transition in the effective $XY$ model (6) is
considerably lower than the temperature at which the local superfluid
density vanishes according to Eq. (2).  A comparison of the phase
diagram
obtained in this way with the results of Monte Carlo calculations (see
below and Fig. \ref{fig1}) shows that they are in fair quantitative
agreement.

\section{QUANTUM MONTE CARLO METHOD.  MEASURABLE
QUANTITIES}

The Trotter discretization procedure makes it possible to estimate all the
thermodynamic averages of the observables of a $D$-dimensional quantum
system in terms of a classical $D+1$-dimensional system, where the
product of the matrix elements (calculated approximately) of the high
temperature density matrix serves as the Boltzmann weight of the
configurations of the corresponding classical system.  For studying the
properties of the model (1), we shall use the ``checkerboard version'' of
the
quantum Monte Carlo method.  (A detailed discussion of the
discretization procedure
and the organization of the Monte Carlo step during the simulation
of
systems of lattice bosons in a large canonical ensemble is given
elsewhere.\cite{28})  In this method, the degrees of freedom of the
discretized system are the occupation numbers $\{ n_i^p \}$ of the sites
of
the $N \times N \times 4P$ three-dimensional lattice formed by
$4P$-fold
multiplication of the initial $N \times N$ lattice along the imaginary
time
axis.  The number of subdivisions $P$ was chosen so that the parameter
$\epsilon = q^2 / P^2 \Theta ^2$, which characterizes the discretization
error, would be less than 0.06.

The density $\nu _s$ of the superfluid component was calculated at each
computational point of the phase diagram,  whose position is specified
by the parameters $\{ \sqrt{U/t}, k_B T/t \}$ and the chemical potential
$\mu /U$.  To find
this quantity, we used both the fluctuations in the topological winding
number\cite{20,28} and the correlation function of the paramagnetic
current.\cite{29}  We found that when the average particle density
at
the boundary $n_0<2$ and $q>2$, the statistical errors in the second
method were considerably higher than the errors in determining the
superfluid density from the fluctuations in the winding number,
rendering it unsuitable.

We also measured $n_0$, which is controlled by the chemical potential
of
the system, and the compressibility modulus $\kappa $, which is
defined as
\begin{equation}
\kappa = k_b T  \partial n_0 / \partial \mu =
\frac{1}{4P N^2}
\left< \sum\limits_{p=0}^{4P-1}\sum\limits_{i} \left( n_{i}^p \right)^2
\right> - \left( n_0 \right)^2
\end{equation}

It turned out to be convenient to make the measurements with
$U/t= {\rm const}$
and $k_BT/t= {\rm const}$ and variation of the chemical potential
$\mu /U$. As
an example, in Fig. \ref{fig2}b the line along which the system moves
for
$\{ U/t,k_BT/t \} =\{3.5, 0.8 \}$ is plotted in the coordinates $\{ q ,
\Theta \}$.
The
location of the system on this line for a given value of the chemical
potential can be determined by measuring the average number $n_0$ of
particles per granule.  In addition, we performed a number of calculations for
fixed $n_0$ (in a canonical ensemble).  It might be expected that for the
same particle density $n_0$ the results of the simulation would be
independent of the choice of the ensemble for sufficiently large systems.
We
tested this assumption and found for a system of dimension
$N\times N = 6 \times 6$ that the difference in the measured quantities
is less than 10\% in the region of variation of the
control parameters of interest to us.  Therefore, in analyzing the results,
data obtained by assigning the density $n_0$ (within a canonical
distribution) and by assigning the chemical potential (which corresponds
to using a grand canonical ensemble) can be used simultaneously.

\begin{figure}
\epsfxsize = 14truecm
\epsffile{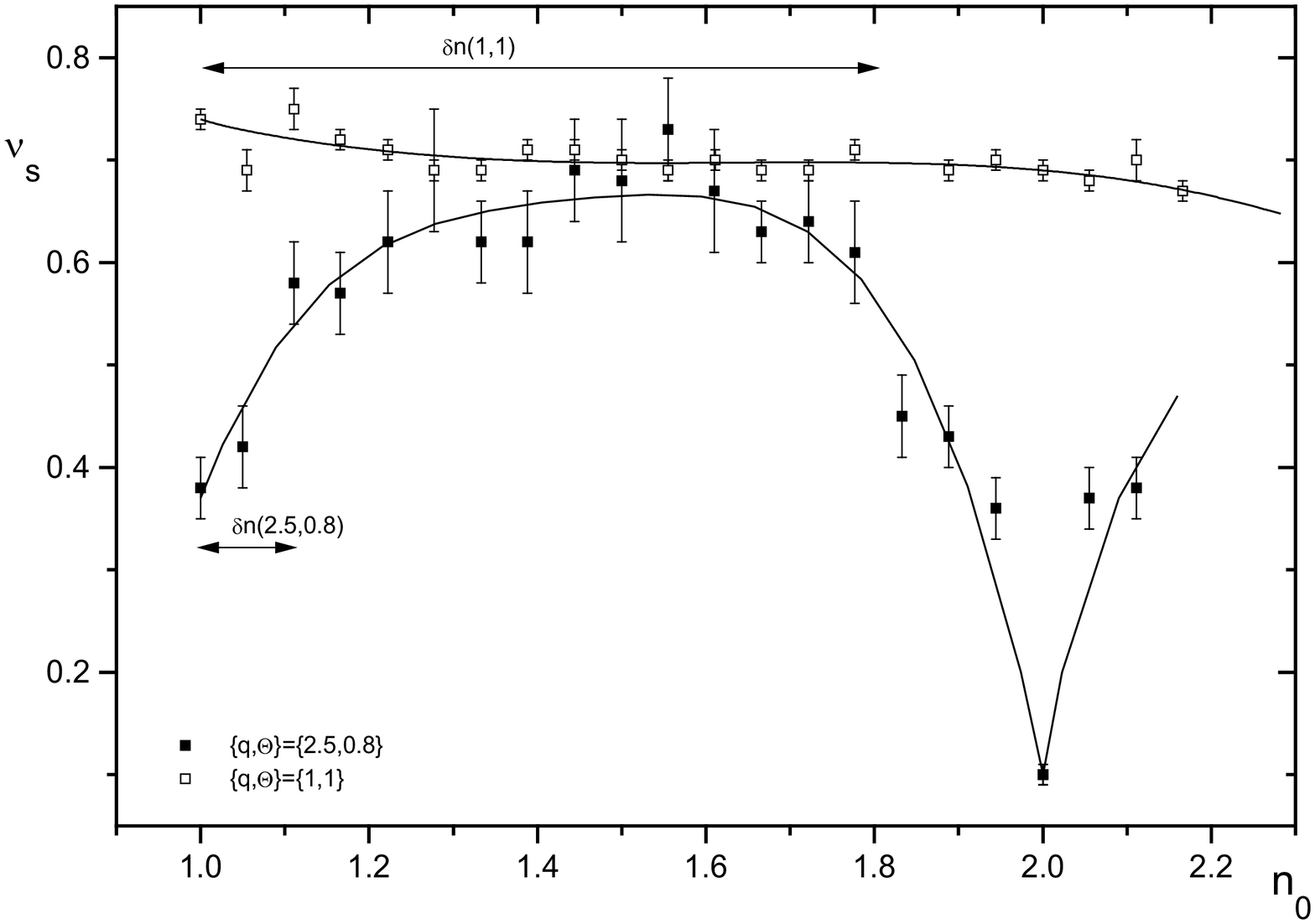}

\caption[]{Plots of the fraction $\nu _s$ of the superfluid component as a
function of the average occupation number $n_0$.  The solid curves
were
obtained by interpolation of the data with a fourth-order polynomial.
The
lengths of the horizontal arrows correspond to estimates of $\delta n (q,
\Theta ) = \kappa (q, \Theta ) | _{n_0=1}$.  See Eq. (8).}
\label{fig3}
\end{figure}

\section{DESCRIPTION AND DISCUSSION OF RESULTS}

We first consider the computational results for $q<1.5$, where the
interparticle interaction is weak and mean-field theory (see Figs.
\ref{fig1}a
and \ref{fig2}a) predicts a monotonic dependence of the phase-transition
temperature on the particle density $n_0$.  Figure \ref{fig3} shows the
calculated density of the superfluid component as a function of
average occupation number, $\nu _s (n_0)$, for $q=1$ and $\Theta =1$
(unfilled symbols).  As the occupation number increases, $\nu _s$
approaches a
constant equal to the helicity modulus $\gamma (q, \Theta )$ in the
$2+1$-dimensional (quantum) $XY$ model.\cite{30,31}  We found
previously\cite{23} that in this region of the phase diagram, this limit,
which corresponds to a small contribution of the quantum fluctuations to
the modulus of the order parameter, is approached already when
$n_0=4$--5.  The monotonicity of $\nu _s(n_0)$ suggests that the results
of experiments on a system of isolated granules will not differ greatly
from those on a system of granules on a substrate with an applied
potential.  It turned out that the system behaves similarly up to
$q \simeq 2.3$.
As the quantum parameter $q$ is raised further and the temperature
$\Theta $ is lowered, $\nu _s (n_0)$ ceases to be a
monotonic
function of the average occupation number $n_0$.  Characteristic
oscillations in $\nu _s(n_0)$ with minima at integer values of $n_0=k$
are noticeable in Fig. \ref{fig3} ($\{ q, \Theta \}= \{ 2.75, 0.5 \}$, filled
symbols).
For sufficiently high boson densities ($n_0 > 7$ in the region $\{ q,
\Theta
\}= \{ 2.5, 0.5 \}$; see Ref. \onlinecite{23}) there is a transition to the
quasiclassical
limit and the density of the superconducting component $\nu _s (n_0)$
becomes a periodic function of the average occupation number with a
period of unity.\cite{14}

\begin{figure}
\epsfxsize = 14truecm
\epsffile{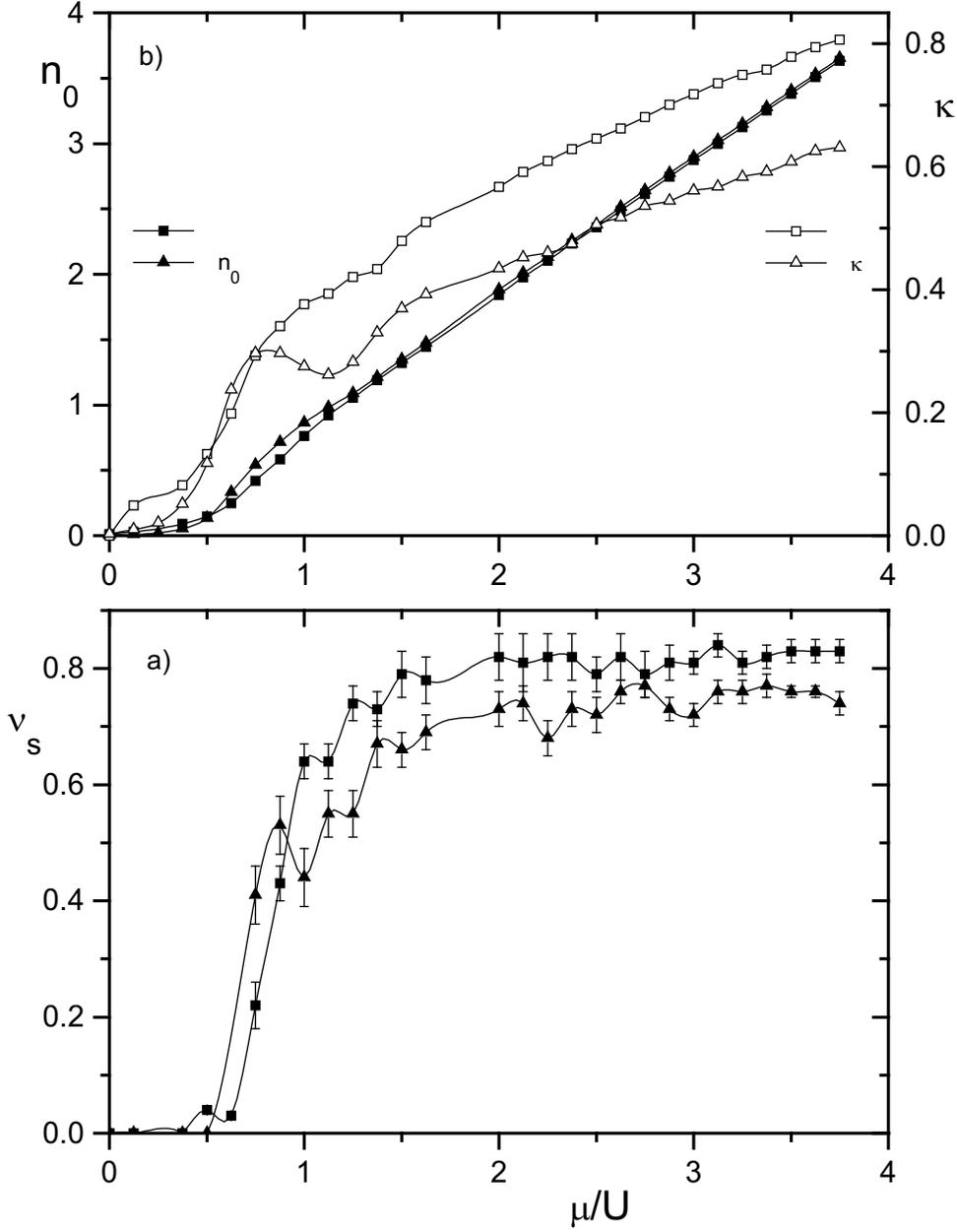}

\caption[]{(a) The superfluid density $\nu _s$ as a function of the
chemical
potential $\mu /U$.  (b) Average particle density $n_0$ (filled symbols)
and
compressibility modulus $\kappa $ (unfilled symbols) as functions of
the
chemical potential $\mu /U$: squares --- $\{U/t, k_B T/t\}= \{2.0, 0.8\}$,
triangles --- $\{U/t , k_B T/t\}= \{2.5, 0.8\}$.  Spline interpolations
are shown
for visual convenience.  When not indicated the statistical errors
are smaller than the sizes of the corresponding symbols.}
\label{fig4}
\end{figure}

Figure \ref{fig4}a shows plots of the superfluid density as a function of
the chemical potential of the system, which is proportional to the voltage
applied to the substrate.  The squares refer to a system with
$\sqrt{U/t}=2$, $k_BT/t=0.8$ and the triangles to $\sqrt{U/t}=2.5$,
$k_BT/t=0.8$.  The feature at $n_0 \simeq 1$ [a dip on the plots of
$\nu _s (\mu /U)$ and $\kappa (\mu /U)$ for at $\mu /U \simeq 1.2$] is
poorly seen for $q \simeq 2$ (squares in Fig. \ref{fig4}) but becomes
clearly evident for $q \simeq 2.5$ (triangles).  The figure confirms that
the case of commensurate populations corresponds to lower densities of
the superfluid component, i.e., increasing the deviation of the average
boson density from integer values of $n_0=k$ leads to spreading of the
phase diagrams for the model (1) in the $\{ q, \Theta \}$ plane.

\begin{figure}
\epsfxsize = 14truecm
\epsffile{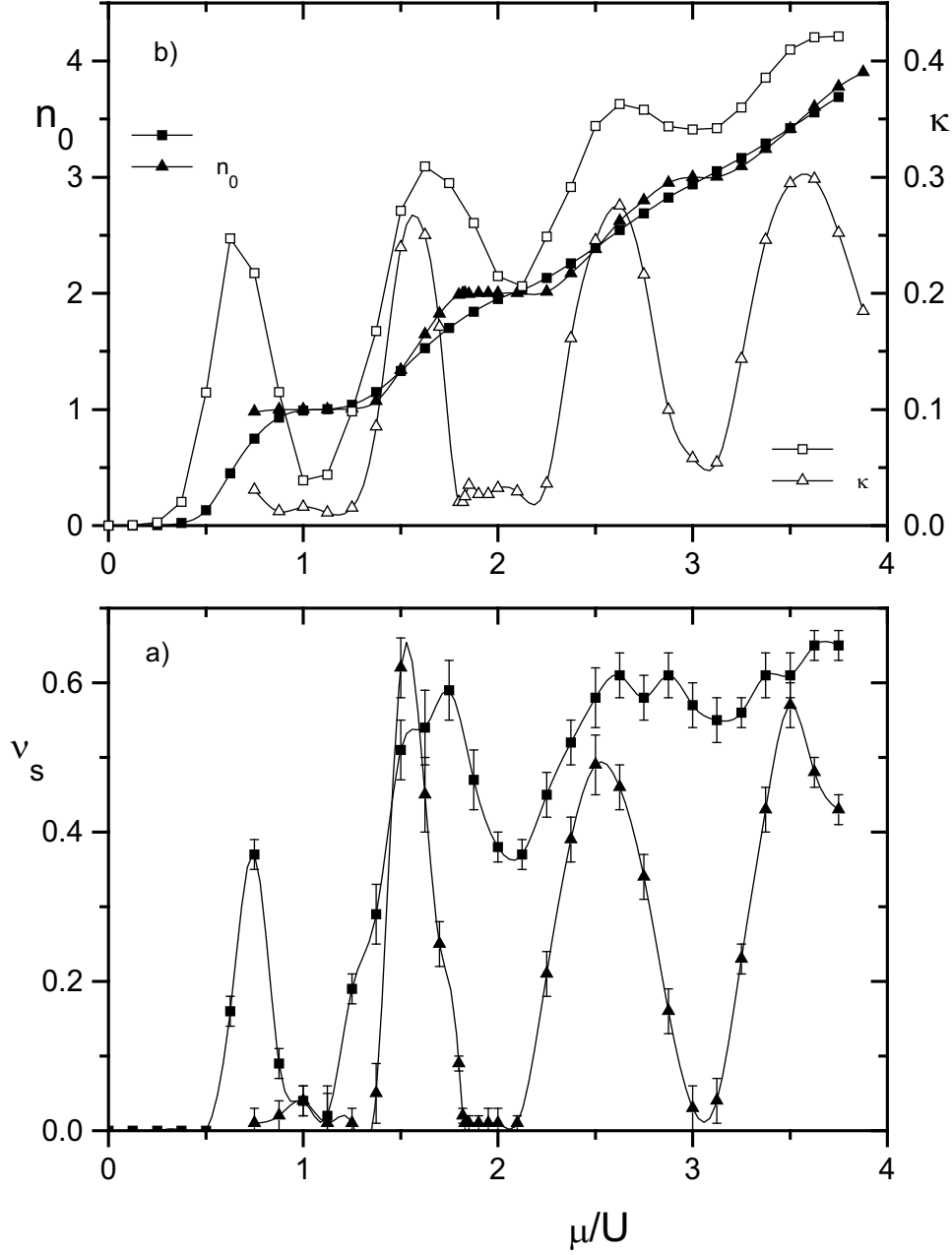}

\caption[]{(a) The superfluid density $\nu _s$ as a function of the
chemical
potential $\mu /U$.  (b) Average particle density $n_0$ (filled symbols)
and
compressibility modulus $\kappa$ (unfilled symbols) as functions of
the
chemical potential $\mu /U$: squares --- $\{U/t , k_B T/t\}= \{3.5,
0.8\}$,
triangles --- $\{U/t , k_B T/t\}= \{5.0, 0.8\}$.}
\label{fig5}
\end{figure}

As the quantum parameter $q$ is increased further, the differences in
the
properties of the system for integer and noninteger boson densities
become increasingly more significant.  In fact, based on the
results
from mean-field calculations (see Fig. \ref{fig1}a), we can assume that
there is a
value of $U/t$ for which the $\sqrt{U/t}={\rm const}$ line intersects
the region
of the disordered state with $n_0 \simeq 1$.  Further increases in the
interaction constant $U$ should lead to the possibility of an intersection
with the disordered region having $n_0 \simeq 2$, etc.  The
computational
results shown in Fig. \ref{fig5} confirm this assumption.  Note that for
$\sqrt{U/t}=5.0$ (triangles in Fig. \ref{fig5}) and integer values of
$\mu /U$, changes in the chemical potential lead to essentially no
change
in the average number of particles at the sites in the system.  This
feature, which is characteristic of an insulator, can also be seen in Fig.
\ref{fig5}b, which shows a plot of the compressibility modulus
$\kappa$ as a function of the chemical potential $\mu /U$.

At $T=0$ a lattice boson system without disorder undergoes a
superconductor-insulator phase transition.\cite{17}  As for the
system being studied here, one can assume that as the temperature is
raised, the domain of existence of the Mott insulator decreases and
shifts toward larger $U/t$, so that, for example, on moving along
the
$\mu /U =k$ line the superfluid phase is replaced by the normal
(metallic)
phase.  Further increases in the interaction constant, under which the
magnitude of the Mott dielectric gap increases so much that thermal
excitations become unimportant, lead to crossover formation of the
insulator state.\cite{32}  An estimate of the location of the boundary of
the insulator state for $n_0=1$, based on our calculations, is shown in
Fig.
\ref{fig1}b.  Several constant-density ($n_0={\rm const}$) contours are
also shown for
comparison.  A more detailed study of crossover-type metal-insulator
transitions requires that systems of substantially larger size be
examined.\cite{24,33,34}

An analysis of Figs. \ref{fig1}--\ref{fig5} shows that interesting effects
caused by
incommensurate populating of the sites in the system occur only in quite
strongly interacting systems and at sufficiently low temperatures.  For
the boson Hubbard model (1), the domain in which they exist can be
estimated by the inequalities $q>2.3$ and $\Theta <0.7$.  Quantitatively,
the magnitude of the deviation $\delta n$ of the average particle density
from an integer value of $n_0=k$ at which a significant change in the
system properties will be observed can be expressed in terms of the
compressibility modulus:
\begin{equation}
\delta n \approx
\left. \frac{\partial n_0}{\partial \mu} \right|_{n_0=k} k_b T =
\left. \kappa \right|_{n_0=k}
\end{equation}
It is known that the compressibility modulus [which is inversely
proportional to
the phase fluctuations in the $2+1$-dimensional $XY$ model; see Eq.
(7)]
falls off substantially as the quantum parameter $q$ becomes greater and
the temperature $\Theta $ is reduced.  We found that the following
estimates hold for the Hubbard model (1) when $n_0=1$:
$\delta n \approx 1.2$
for $\{q , \Theta \} \approx \{ 0.5 , 1 \}$, $\delta n \approx 0.3$ for
$\{q , \Theta \} \approx \{ 2.5 , 0.8 \}$, and $\delta n \approx 0.06$ for
$\{q , \Theta\} \approx \{ 2.75 , 0.5 \}$.  As an
illustration, Fig. \ref{fig3} shows the values of $\delta n (q, \Theta )$
found for the points $\{q , \Theta \} = \{ 1 , 1 \}$ (unfilled symbols) and
$\{q , \Theta \} = \{ 2.5, 0.8 \}$ (filled symbols).  The figure
demonstrates
the fair agreement between the theoretical estimate (8) and the Monte
Carlo calculations.

The above results yield the phase diagram of the system shown in Fig.
\ref{fig1}b (in the coordinates $\{ \sqrt{U/t}, \mu /T \}$ for
$k_BT/t=0.8$) and in Fig. \ref{fig2}b (in the coordinates
$\{q, \Theta \}$).  The location of the boundary of the ordered
superconducting state was estimated from the universal jump in the
superfluid density\cite{30} and from the location of the peak in its
temperature derivative.\cite{35}

\begin{figure}
\epsfxsize = 14truecm
\epsffile{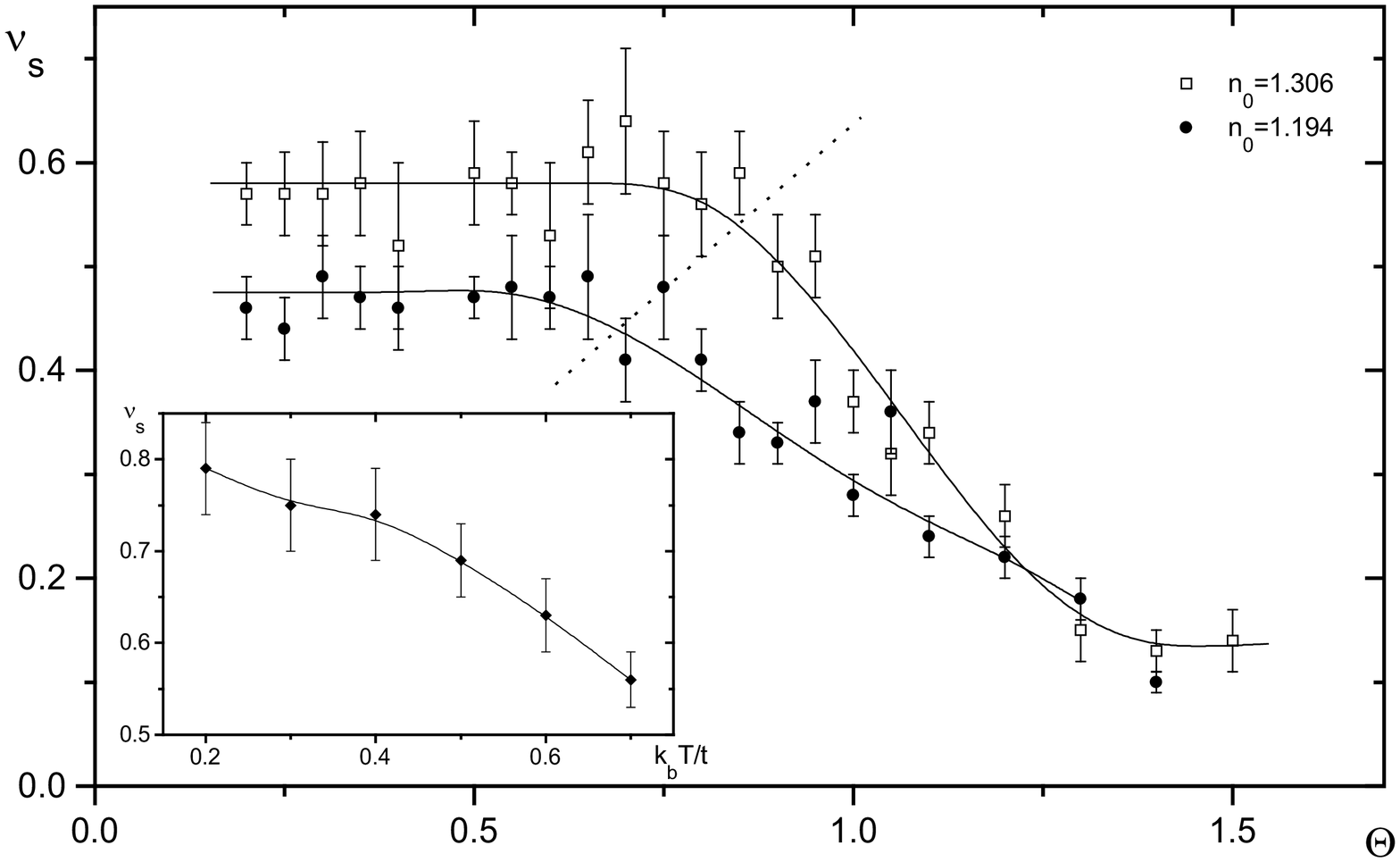}

\caption[]{Superfluid density of the system as a function of
$\Theta = k_BT/tn_0$ for $q=3.0$; unfilled symbols --- $n_0=1.306$;
filled symbols --- $n_0=1.194$.  The dotted line indicates the straight
line $\nu _s = 2 \Theta/\pi$.  The inset shows the calculated superfluid
density of the system as a function of $k_BT/t$ for $\sqrt{U/t}=2.5$
and $\mu /U=0.75$.}
\label{fig6}
\end{figure}

Quantum phase transitions (at $T=0$) in the two-dimensional Hubbard
model (1) are determined by the critical properties of the corresponding
effective three-dimensional system.\cite{17,32}  At finite temperatures
a
Josephson array will display Kosterlitz--Thouless critical behavior on
some $\Theta _c (\mu /U, q)$ curve.  We now try to evaluate the effect
of
quantum fluctuations on the temperature of this transition, assuming it
to
be rather low (see below).  There is interest in two cases: (a) the
system
has a commensurate particle density $n_0=k$ and at some temperature
$\Theta _c(q)$ it undergoes a transition from the superconducting to the
normal state, and (b) a phase transition takes place at $\Theta _c(n_0)$
because of a change in the density $n_0$.

For the region near the point $q_c^{XY} \approx 2.5$ of the quantum
transition in the $2+1$-dimensional $XY$ model (see Fig. \ref{fig2}b),
the
temperature $\Theta _c(q)$ of the topological Kosterlitz--Thouless phase
transition has been estimated\cite{24} as $\Theta _c(q) \sim | q -
q_c^{XY}|^{\zeta _1}$ with $\zeta _1 \approx 0.67$.  This estimate was
derived under the assumption that the system temperature is less than
the
temperature of the $2D \rightarrow 3D$ crossover; i.e., $\Theta \leq
\Theta _{3D}$, where $\Theta _{3D} \sim | q - q_c^{XY} |^{\nu }$
with
$\nu \approx \zeta _1$.  This makes it possible to estimate the density of
the superfluid component, $\nu _s (q, \Theta )$, which determines the
phase-transition temperature $\Theta _c (q)$, as $\nu _s (q, \Theta
) \approx \nu _s (q, 0)$.  Evidently, similar arguments apply in the
neighborhood of the points $q_c^H | _{n_0=k}$ of the quantum phase
transitions in the Hubbard model (1) for integer populating of the
array
granules, i.e., $n_0 =k$ ($q_c^H|_{n_0=1} \approx 2.8$; see
Ref. \onlinecite{20}). Thus, we may expect that
\begin{equation}
\Theta^c(q;k) \sim \left|\left. q^c_H \right|_{n_0=k} - q \right|^{0.67}
\end{equation}

Similar arguments can also be applied to the case of incommensurate
boson densities, $n_0 \neq k$.  It has been shown\cite{17,20} that
$\nu _s \sim | n_0 -k|^{\zeta _2 }$ with $\zeta _2 \approx 1.0$ for
$q>q_c^H$.  Thus, the following relation holds for the temperature
$\Theta_c(n_0)$ of the topological Kosterlitz--Thouless phase transition:
\begin{equation}
\Theta^c(n_0) \sim |n - k|^{1.0}
\end{equation}
Our quantum calculations are in qualitative agreement with
the predictions of Eqs. (9) and (10), but it is difficult to confirm their
validity with sufficient accuracy because of the large errors in
determining the position of the quantum phase-transition line $\mu
(U/t; T=0)$.

There is great interest in the question of the existence of reentrant
superconductivity, for which, within some range of variation of the
quantum parameter $q$, disorder sets in not only as the temperature
$\Theta $ is raised, but also as it is lowered.  The existence of reentrant
effects
has been predicted a number of times within the quantum $XY$ model
(see Ref.
\onlinecite{13} and the literature cited therein), but as far as we know,
computer
simulation cannot unequivocally confirm\cite{31,36} or refute\cite{34}
the existence of
this phenomenon.  The earlier numerical calculations of the Hubbard
model did not reveal low-temperature instability or reentrant
superconductivity.\cite{20,23}  For this paper we studied the low
temperature
region $q \approx 2.5$, $\Theta < 0.5$ of the boson Hubbard model
both for
different noninteger occupation numbers and for fixed values of the
chemical potential.  The results, shown in Fig. \ref{fig6}, indicate a lack
of reentrant superconductivity effects, at least within the range of
variation of the control parameters studied here.  Special attention
was devoted to the region $\{ \sqrt{U/t}, \mu /U \} \approx \{ 2.5, 0.8
\}$,
within which low-temperature disorder has been predicted.\cite{19}  The
numerical simulations clearly showed that there were no such effects
within this region.  In addition, our calculation of the boundary of the
ordered state using mean-field theory (see Sec. 2) also
disagrees with these predictions.  Note that Eq. (2), which specifies the
boundary of the ordered state, is more accurate in this method than is
the equation
used in the paper just cited inasmuch as its derivation did not rely on the
assumption that $n_0 \gg 1$ [the latter is valid for $\mu /U \gg
1$ (Ref. \onlinecite{19})].

To conclude, we have
analyzed the effect of quantum fluctuations in the phase and modulus of
the superconducting or superfluid order parameter on the character
of
the ordering in two-dimensional mesoscopic Josephson and granular
systems within a lattice boson Hubbard model.  Quantum
Monte Carlo calculations have been used to show that the way the
system
properties change as a result of modulation of the average occupation
number of the array elements by the chemical potential (the substrate
potential) is determined by the parameter $q=\sqrt{U/tn_0}$ (i.e., the
ratio of the characteristic Coulomb energy of a granule to the Josephson
tunneling energy).  For $q<1.5$, which is the quasiclassical region for
the
quantum $XY$ model and the region of strong fluctuations in the
modulus
of the order parameter for the Hubbard model (1), the system
properties
are insensitive to the average number of particles in the granules.  In the
region where there are significant quantum fluctuations in the order
parameter ($q>2$, $\Theta < 0.8$), we have found that the state of the
system depends (more distinctly at lower temperatures) on the average
number of particles in it.

This work was supported by grants from the Russian Foundation for
Basic
Research and the Program on the Physics of Solid-State Nanostructures.

Translated by D. H. McNeill


\begin{references}
\bibitem[*)]{A}E-mail: lozovik@isan.troitsk.ru
\bibitem[1)]{1)}Two-dimensional Josephson arrays (of
mesoscopic elements) with superfluid helium can, in principle, be formed
by creating the corresponding ``picture'' in cesium on the substrate
(since cesium is not wetted by helium).\cite{1}
\bibitem[2)]{2)}The interference of two Bose condensates has recently
been studied.\cite{11}
\bibitem{1}F. J. Nacker and J. Dupont-Roc, Phys. Rev. Lett. {\bf 67},
2966
(1991).
\bibitem{2}J. D. Reppy, J. Low Temp. Phys. {\bf 67}, 207 (1992).
\bibitem{3}H. S. J. van der Zant, F. C. Fritschy, J. E. Mooij {\it et al.},
Phys. Rev.
Lett. {\bf 69}, 2971 (1992); J. E. Mooij, R. Fazio, G. Sch\"{o}n
{\it et al.}, Phys. Rev. Lett. {\bf 65}, 645 (1990).
\bibitem{4}V. G. Gantmakher, V. M. Teplinski\u{i}, and V. N. Zverev,
Pis'ma
Zh. \'{E}ksp. Teor. Fiz. {\bf 62}, 873 (1995) [JETP Lett. {\bf 62}, 887
(1995)].
\bibitem{5}A. F. Hebard and M. A. Paalanen, Phys. Rev. Lett. {\bf 65},
927
(1990).
\bibitem{6}A. L. Dobryakov, Yu. E. Lozovik, A. A. Puretzky {\it et al.},
Appl. Phys. A {\bf 54}, 100 (1992).
\bibitem{7}Yu. M. Mucharsky, A. Loshak, K. Schwab {\it et al.},
Czech. J. Phys.
{\bf 46}, 115 (1996); S. V. Pereverzev, A. Loshak, S. Backhaus
{\it et al.}, Nature {\bf 388}, 449 (1997).
\bibitem{8}M. N. Anderson, J. R. Ensher, M. R. Mathews {\it et al.},
Science {\bf 269}, 198 (1995).
\bibitem{9}C. C. Bradley, C. A. Sackoff, J. J. Tollett {\it et al.},
Phys. Rev. Lett. {\bf 75}, 1687 (1995).
\bibitem{10}K. B. Davis, M.-O. Mewes, M. R. Andrew {\it et al.},
Phys. Rev. Lett. {\bf 75}, 3969 (1995).
\bibitem{11}M. R. Andrews, C. G. Towsend, J.-J. Miesner {\it et al.},
Science {\bf 275}, 637 (1997).
\bibitem{12}Yu. E. Lozovik, submitted to Physica E (Amsterdam); Yu.
E.
Lozovik and
O. L.
Berman, Zh. \'{E}ksp. Teor. Fiz. {\bf 111}, 1879 (1997) [JETP {\bf 84},
1027
(1997)]; Yu. E. Lozovik, O. L. Berman, and V. G. Tsvetus, Pis'ma Zh.
\'Eksp.
Teor. Fiz. {\bf 66}, 332 (1997) [JETP Lett. {\bf 66}, 355 (1997)].
\bibitem{13}B. J. Kim and M. Y. Choi, Phys. Rev. B {\bf 52}, 3624
(1995); B.
J. Kim, J. Kim, M. Y. Choi {\it et al.}, Phys. Rev. B {\bf 56}, 395
(1997).
\bibitem{14}C. Bruder, R. Fazio, A. P. Kampf {\it et al.}, Phys. Scr. T
{\bf
42}, 159 (1992).
\bibitem{15}I. E. Dzyaloshinskii, E. M. Lifshitz, and L. P. Pitaevskii,
Adv.
Phys. {\bf 10}, 165 (1961).
\bibitem{16}G. T. Zimanyi, P. A. Crowell, R. T. Scalettar {\it et al.},
Phys. Rev. B {\bf 50}, 6515 (1994).
\bibitem{17}M. P. A. Fisher and G. Grinstein, Phys. Rev. Lett. {\bf 60},
208
(1988); M. P. A. Fisher, P. B. Weichman, G. Grinstein, and D. S. Fisher,
Phys.
Rev. B {\bf 40}, 546 (1989); M. P. A. Fisher, G. Grinstein, and S. M.
Girvin,
Phys. Rev. Lett. {\bf 64}, 587 (1990).
\bibitem{18}M. C. Cha, M. P. A. Fisher, S. M. Girvin {\it et al.},
Phys. Rev. B {\bf 44}, 6883 (1991).
\bibitem{19}A. P. Kampf and G. T. Zimanyi, Phys. Rev. B {\bf 47}, 279
(1993).
\bibitem{20}W. Knauth, N. Trivedi, and D. Ceperley, Phys. Rev. Lett.
{\bf
67}, 2703 (1991); W. Krauth and N. Trivedi, Europhys. Lett. {\bf 14},
627
(1991).
\bibitem{21}V. A. Kashurnikov, A. V. Krasavin, and B. V. Svistunov,
Pis'ma
Zh. \'{E}ksp. Teor. Fiz. {\bf 64}, 92 (1996) [JETP Lett. {\bf 64}, 99
(1996)].
\bibitem{22}A. V. Otterlo and K. H. Wagenblast, Phys. Rev. Lett. {\bf
72},
3598 (1994); E. Roddick and D. Stroud, Phys. Rev. B {\bf 51}, 8672
(1995).
\bibitem{23}A. I. Belousov, S. A. Verzakov, and Yu. E. Lozovik, Zh.
\'{E}ksp.
Teor. Fiz. {\bf 113}, 261 (1998) [JETP {\bf 86}, 146 (1998)]; A. I.
Belousov
and Yu. E. Lozovik, Pis'ma Zh. \'{E}ksp. Teor. Fiz. {\bf 66}, 649 (1997)
[JETP
Lett. {\bf 66}, 686 (1997)].
\bibitem{24}S. Doniach, Phys. Rev. B {\bf 24}, 5063 (1981).
\bibitem{25}V. N. Popov, {\it Functional Integrals in Quantum Field
Theory and Statistical Physics}, Reidel, Dordrecht (1983).
\bibitem{26}J. J. Alvarez and C. A. Balseiro, Solid State Commun.
{\bf 98}, 313 (1996).
\bibitem{27}P. Olsson, Phys. Rev. B {\bf 52}, 4511 (1995).
\bibitem{28}A. Blaer and J. Han, Phys. Rev. A {\bf 46}, 3225 (1992).
\bibitem{29}G. G. Batrouni, B. Larson, R. T. Scalettar {\it et al.},
Phys. Rev. B {\bf 48}, 9628 (1993).
\bibitem{30}P. Minnhagen, Rev. Mod. Phys. {\bf 59}, 1001 (1987).
\bibitem{31}M. Jacobs, J. V. Jose, M. A. Novotny {\it et al.},
Phys. Rev. B {\bf 38}, 4562 (1988).
\bibitem{32}S. L. Sondhi, S. M. Girvin, J. P. Carini {\it et al.},
Rev. Mod. Phys. {\bf 69}, 315 (1997).
\bibitem{33}Yu. E. Lozovik and S. G. Akopov, J. Phys. C {\bf 14}, L31
(1981);
S. G. Akopov and Yu. E. Lozovik, J. Phys. C {\bf 15}, 4403 (1982).
\bibitem{34}A. I. Belousov and Yu. E. Lozovik, Solid State Commun.
{\bf 100}, 421 (1996); A. I. Belousov and Yu. E. Lozovik, Fiz. Tverd. Tela
(St. Petersbourg) {\bf 39}, 1513 (1997) [Phys. Solid State {\bf 39}, 1345
(1997)]; S. A. Verzakov and Yu. E. Lozovik, Fiz. Tverd. Tela
(St. Petersbourg) {\bf 39}, 818 (1997) [Phys. Solid State {\bf 39}, 724
(1997)].
\bibitem{35}F. F. Assaad, W. Hanke, and D. J. Scalapino, Phys. Rev. B
{\bf
50}, 12\,835 (1994).
\bibitem{36}D. Marx and P. Nielaba, J. Chem. Phys. {\bf 102}, 4538
(1995).
\end{references}
\end{document}